
%
\input amstex
\documentstyle{amsppt}
%
\chardef\tempcat=\the\catcode`\@
\catcode`\@=11
%
%
\ifx\undefined\linespacing
 \def\linespacing#1{%
  \addto\tenpoint{\normalbaselineskip=#1\normalbaselineskip
     \normalbaselines
     \setbox\strutbox=\hbox{\vrule height.7\normalbaselineskip
       depth.3\normalbaselineskip}%
     \setbox\strutbox@\hbox{\raise.5\normallineskiplimit
       \vbox{\kern-\normallineskiplimit\copy\strutbox}}%
   }%
  \addto\eightpoint{\normalbaselineskip=#1\normalbaselineskip
     \normalbaselines
     \setbox\strutbox=\hbox{\vrule height.7\normalbaselineskip
       depth.3\normalbaselineskip}%
     \setbox\strutbox@\hbox{\raise.5\normallineskiplimit
       \vbox{\kern-\normallineskiplimit\copy\strutbox}}%
   }%
 }
\fi
%
\ifx\undefined\nologo
 \def\nologo{\def\logo@{}}
\fi
%
%
\font@\eightsmc=cmcsc10 at 8.3333333pt
\addto\eightpoint{\font@\eightmsb=msbm10 at 8.3333333pt
\textfont\msbfam=\eightmsb
\font@\eightex=cmex10 at 8.3333333pt
\textfont\thr@@=\eightex}
\addto\tenpoint{\font@\sevenmsa=msam10 at 8.3333333pt
\font@\sevenex=cmex10 at 8.3333333pt
\scriptfont\msafam=\sevenmsa
\scriptfont\thr@@=\sevenex}
\catcode`\@=\tempcat
%
%
\define\[{\left[}
\define\]{\right]}
\define\({\left(}
\define\){\right)}
\define\>{\hfil\ifmmode\mathbreak\else\break\fi}
\define\ao#1{#1{}^\circ}
\define\ah#1{\widehat{#1}}
\define\p(#1,#2){\left\langle#1,#2\right\rangle}
\define\QED{\ifmmode\qed\else{{\unskip\nobreak\hfil
\penalty50\null\nobreak\hfil$\qed$
\parfillskip=0pt\finalhyphendemerits=0\endgraf}\fi}}
%
\hyphenation{Berezin-ian}
\define\Diff{\operatorname{Diff}}
\define\Smbl{\operatorname{Smbl}}
\define\Der{\operatorname{D}}
\define\DER{\operatorname{Der}}
\define\J{\operatorname{\Cal J}}
\define\JI{\operatorname{\J^\infty}}
\define\com{\operatorname{\circ}\limits}
\define\T{\operatorname{T}}
\define\Ber{\operatorname{Ber}}
\define\G{\operatorname{G}}
\define\F{\operatorname{F}}
\define\Cl{\operatorname{Cl}}
\define\im{\operatorname{im}\nolimits}
\define\coker{\operatorname{coker}\nolimits}
\define\Hom{\operatorname{Hom}\nolimits}
\define\Hm{\operatorname{H}\nolimits}
%
%
%
\mag=\magstep 1
\pagewidth{15 true cm}
\pageheight{23 true cm}
\hcorrection{0.6 true cm}
\vcorrection{0.9 true cm}
\linespacing{1.5}
%
%
\NoBlackBoxes
\hfuzz=1pc
\vbadness=10000
\scrollmode
\nopagenumbers
\pageno=1
\TagsOnRight
\nologo
\refstyle{C}
\widestnumber\no{99}
%
%

%
%
\topmatter
\title Lagrangian formalism over graded algebras
\endtitle
\author  Alexander Verbovetsky \endauthor
\rightheadtext{Lagrangian formalism over graded algebras}
\affil
Scuola Internazionale Superiore di Studi Avanzati,\\
Trieste, Italy
\endaffil
\date {July 1994} \enddate
\address
S.I.S.S.A., Via Beirut 2-4, 34013 Trieste, Italy
\endaddress
\email verbovet\@sissa.it \endemail
\keywords Graded module, commutation factor,
adjoint operator, Berezinian, complex of integral forms, Lagrangian
formalism, supermanifold, noncommutative differential calculus
\endkeywords
\subjclass Primary 58E30, 58A50, 58B30;
Secondary 81T60, 83E50, 16W50, 46L87, 14A22 \endsubjclass
\abstract
This paper provides a description of an algebraic setting for the
Lagrangian formalism over graded algebras and is intended as the
necessary first step towards the noncommutative $\Cal C$-spectral
sequence  (variational bicomplex). A noncommutative version of
integration  procedure, the notion of adjoint operator, Green's
formula, the  relation between integral and differential forms,
conservation laws,  Euler operator, Noether's theorem is considered.
\endabstract
\endtopmatter
\newpage
%
%
\document
\head Introduction \endhead
An outstanding progress which has in last years been made by
noncommutative geometry stems out of shifting of the interests away
from the original idea of  geometrizing noncommutative rings using
the language of noncommutative schemes. Now it has become clear that,
bypassing the difficult problem of glueing noncommutative spectra, one
can define directly differential geometric objects on a  hypothetical
``noncommutative space''. This is based on two essential points:  the
possibility of reformulating the classical notions of analysis and
differential geometry in pure algebraic terms, so that differential
calculus  becomes an extension of the language of commutative algebra
(for a very  enlightening discussion see  \cite{1}), and the existence
of several quite important cases in which one is able to go beyond the
commutative case (see, e\.g., \cite{2}). These ideas  have proven to be
very successful and give an impetus to new researches whose  aim is to
transplant the classical tools of analysis and geometry into a
noncommutative setting. An instance of this comes from the current work
on noncommutative ( and first of all supercommutative) theory of
integration, conservation laws, and the Lagrangian formalism (see
\cite{3},  \cite{4}, \cite{5}, \cite{6}, \cite{7}, \cite{8}, \cite{9},
\cite{10}, \cite{11}, \cite{12},  \cite{13}, \cite{14}, \cite{15},
\cite{16}, \cite{17} and  others). A key problem arises here is the
description of integration procedure. The first question is the
following: Given a commutative algebra, how to define the module of
volume forms? An answer to this question must, in particular, give the
possibility to define the Berezin volume forms on a supermanifold in
usual fashion, i\.e., by using the rule of signs. The well-known
peculiarities of the Berezin integration show the character of the
problem and result in the loss of a clear algebraic setting for
integration procedure and related things.

The subject of this paper is an algebraic picture underlying the
Lagrangian formalism and giving a solution of the problems discussed
above. We work here  over an arbitrary graded-commutative (with respect
to a commutation factor) algebra to show that our constructions of
volume forms, adjoint operators, the Euler operator, algebraic Green's
formula, Noether's theorem, etc\. can be extended to such an algebra in
a simple and straightforward manner. On the other hand, the class of
graded-commutative algebras is quite important for its own sake because
it includes supercommutative, colour-commutative (see \cite{18},
\cite{19}, \cite{20}, \cite{21}, \cite{22}, \cite{23}, \cite{24},
\cite{25}   and references therein), and ``quantum'' algebras (see
\cite{26},\cite{20}, \cite{27} and others). We have chosen here not to
accumulate  formulas for specific algebras, but to present a general
scheme, using examples for illustrations only. The applications will be
described separately. Our ultimate goal, which is outside the present
paper, is to develop the super- and non-commutative generalizations of
the $\Cal C$-spectral sequence (variational bicomplex) (see \cite{28},
\cite{29}, \cite{30}, \cite{31}), which is a means for studying all
aspects of the Lagrangian formalism: the inverse problem, the
description of conservation laws, characteristic classes and so on.

The paper is organized as follows. In Section 1 we outline the
necessary definitions and facts from calculus over graded algebras. In
Section 2 the main objects of this paper\thinspace@---\thinspace
adjoint operators and  the Berezinian\footnote{We use this name for the
module of volume forms.}\thinspace@---\thinspace  are defined. We want
to emphasize that these definitions, which are the {\it deus ex
machina\/} from that everything else follows, arose from an interplay
between ideas and constructions of work \cite{32}, where the Berezin
forms have been explained  in terms of $\Cal D$-modules, and works
\cite{33} and \cite{29, part I},  where structure of Lagrangian
formalism for a smooth commutative algebra has been clarified. In
Section 3 we consider the Spencer complexes, algebraic Green's formula,
and related staff. The main difference from non-graded case (see
\cite{29} is the appearance of a new complex dual to the de~Rham
complex: complex of integral forms (the name borrowed from the
supergeometry). The Lagrangian formalism, theory of conservation laws,
and the Noether theorem are developed in Section 4. In the Appendix we
briefly describe, in a pure algebraic manner, the formalism of right
connections (see \cite{34}, \cite{35}), which is  closely related with
our subject.
\head 1. A Sketch of Differential Calculus over Graded Algebras
\endhead
\subhead 1.1 \endsubhead
We start with definitions of graded objects (see, e\.g., \cite{36}).

Let $G$ be an Abelian group written additively, which will serve as a
grading group, and $K$ a commutative ring with unit.

Denote by $K^*$ the multiplicative group of invertible elements of K.
Let us  fix a {\it commutation factor\/} $\{\cdot\,,\cdot\}\:G\times
G\to K^*$, $g_1\times{}g_2\mapsto\{g_1,g_2\}$, i\.e., a map satisfying
two properties:
\roster
\item"{1.}" $\{g_1,g_2\}^{-1}=\{g_2,g_1\}$
\item"{2.}" $\{g_1+g_2,g_3\}=\{g_1,g_3\}\{g_2,g_3\}$
\endroster
We have also $\{g_1,g_2+g_3\}=\{g_1,g_2\}\{g_1,g_3\}$ as a consequence
of the definition.
\example{Example 1} Let $G=\Bbb Z^n=\Bbb Z\oplus\ldots\oplus\Bbb Z$.
Then it is easily shown that any commutation factor has the form:
$$\{g,h\}=\prod_{i=1}^n q_i^{g_ih_i}\cdot\prod_{i<j}
q_{ij}^{\(g_ih_j-h_ig_j\)},$$ where $g=\(g_1,\cdots,g_n\),
h=\(h_1,\cdots,h_n\) \in\Bbb Z^n, q_i=\pm 1,  q_{ij}\in K^*$
\endexample
\example{Example 2} In particular, if $G=\Bbb Z$ there exist two
commutation factors: the trivial one $\{g_1,g_2\}=1$ and the standard
superfactor $\{g_1,g_2\}=\(-1\)^{g_1g_2}$.
\endexample
\example{Example 3} If $G=\Bbb Z^n$ and $K=\Bbb C$ then
$\{g,h\}=\prod_{i=1}^n \(\pm1\)^{g_ih_i}\cdot e^{\theta\(g,h\)}$, where
$\theta\:G\times G\to\Bbb C$ is an antisymmetric bilinear form.
\endexample
Suppose $A=\oplus_{g\in G}A_g$ is a $G$-graded associative $K$-algebra
with unit. $A$ is called {\it graded commutative\/} if
$$ab=\{a,b\}ba\qquad\forall  a,b\in A.$$ For this type of notation we
always assume that $a$ and $b$ are homogeneous and that the symbol of
graded object used as argument of the  commutative factor denotes the
grading of this object.
\example{Example 4} A commutative algebra (graded or not) is graded
commutative with respect to the trivial commutation factor
$\{g_1,g_2\}=1$.
\endexample
\example{Example 5} The algebra $C^\infty\(M\)$ of smooth functions on
a supermanifolds $M$ is graded commutative with respect to the
superfactor $\{g_1,g_2\}=\(-1\)^{g_1g_2}, G=\Bbb Z_2$.
\endexample
\example{Example 6. Quantum superplane (\cite{37}, \cite{2})} Let $K$
be a  field  and $K^n$ the $n$-dimensional coordinate vector space over
$K$. Picking up an arbitrary commutation factor over $\Bbb Z^n$ (see
Examples 1--3), the quadratic algebra $A=\T\(K^n\)/R$, where
$\T\(K^n\)$ is the tensor  algebra over $K^n$ with the natural $\Bbb
Z^n$-grading and $R$ is the ideal in  $\T\(K^n\)$ generated by the
elements $x_1\otimes x_2-\{x_1,x_2\}x_2\otimes x_1$, $x_1,x_2\in K^n$,
is called the  algebra of polynomial functions on the quantum
superplane $\Bbb A_q$.
\endexample
\example{Example 7. Non-commutative torus (\cite{38}, \cite{2})} Let
$K=\Bbb C$, $G=\Bbb Z^n$, and $A$ be the space of all complex valued
functions on $\Bbb Z^n$ that decay faster than any polynomial. $A$ has
the natural $\Bbb Z^n$-grading. Denote by $e\(k\)$ the function which
is equal to 1 at  $k\in\Bbb Z^n$ and zero at all other points (a basic
harmonic on a non-commutative torus). Then any element of $A$ can be
written as $\sum_k a_ke\(k\)$. Given a skew bilinear form $\theta\:\Bbb
Z\times\Bbb Z\to\Bbb R$, we define a multiplication on $A$ by setting
$e\(k\)e\(l\)=e^{2\pi{}i\theta\(k,l\)}e\(k+l\)$. It is straightforward
to check that the algebra $A$, called the algebra of (smooth) functions
on the non-commutative torus $\Bbb T_{\theta}$, is graded commutative
with respect to the commutation factor from Example 3.
\endexample
\remark{Remark} In \cite{39}, \cite{40} Lychagin has shown that one can
also  define commutation factors over a non-commutative group $G$ and
introduce all notions  of the $G$-graded differential calculus in these
circumstances.
\endremark
In this article we work in the category $\Cal M od_A$ of all $G$-graded
modules over a graded commutative algebra $A$. Clearly any left module
$P$ can be transformed canonically into a right module, $pa=\{p,a\}ap$
for $a\in A$, $p\in P$, so that we may consider $P$ as a bimodule.
\remark{Remark} The category $\Cal M od_A$ is a {\it closed tensor
category\/} (see, e\.g.,  \cite{2}) with respect to the commutativity
constraint $$S_{P,Q}\:P\otimes  Q\to Q\otimes P,\qquad p\otimes
q\mapsto \{p,q\}q\otimes p.$$ A generalization of our constructions to
an arbitrary Abelian closed tensor category appears to be very
interesting. Significant results along this line, dealing with basic
concepts  of the differential calculus, may be found in \cite{41},
\cite{42}.
\endremark
\subhead 1.2 \endsubhead
We now introduce basic objects of the differential calculus(for details
see  \cite{43}, \cite{44}, \cite{45}, \cite{46}).

Let $\Delta\in\Hom_K\(P,Q\)$ be a $K$-homomorphism, $P$ and $Q$ being
$A$-modules. For every element $a\in A$ define a $K$-homomorphism
$\delta_a\(\Delta\)\:P\to Q$ by setting
$\delta_a\(\Delta\)\(p\)=\{a,\Delta\}\Delta\(ap\)-a\Delta\(p\)$, $p\in
P$. Obviously,
$\delta_a\com\delta_b=\delta_b\com\delta_a\quad\forall{}a,b\in A$. Put
$\delta_{a_0,\ldots,a_k}=\delta_{a_0}\com\ldots\com\delta_{a_k}$.
\definition{Definition} A $K$-homomorphism $\Delta\in\Hom_K\(P,Q\)$ is
called a  {\it differential operator\/} (d.o.) of order $\leqslant{}k$,
if for all  $a_0,\ldots,a_k\in A$ we have
$\delta_{a_0,\ldots,a_k}\(\Delta\)=0$.
\enddefinition
The set of all d.o.'s of order $\leqslant{}k$, from $P$ to $Q$, may be
endowed with two $A$-module structures by putting $a\Delta=a\com\Delta$
or  $a\Delta=\{a,\Delta\}\Delta \com{}a$, where $a\in A$ is understood
as the operator of multiplication by $a$. The  modules that arise in
this way are denoted by $\Diff_k\(P,Q\)$ and $\Diff_k^+\(P,Q\)$
respectively. Clearly,
$\Diff_k^{\(+\)}\(P,Q\)\subset\Diff_l^{\(+\)}\(P,Q\)$ for
$k\leqslant{}l$, so that we may consider the union
$\Diff^{\(+\)}\(P,Q\)= \bigcup_{k\geqslant{}0}\Diff_k^{\(+\)}\(P,Q\)$.
\proclaim{Proposition} {\it \roster
\item If $\Delta_1\in \Diff_k\(P,Q\)$, $\Delta_2\in \Diff_l\(Q,R\)$,
$P,Q,R$ are  $A$-modules, then
$\Delta_2\com\Delta_1\in\Diff_{k+l}\(P,R\)$.
\item The maps $\Diff_k\(P,Q\)\to\Diff_k^+\(P,Q\)$ and
$\Diff_k^+\(P,Q\)\to \Diff_k\(P,Q\)$ generated by the identity map are
d.o.'s of order $\leqslant{}k$.
\item There exists a canonical isomorphism $$\Diff_k^+\(P,Q\)\to
\Hom_A\(P,\Diff_k^+\(Q\)\),$$ where $\Diff_k^+\(Q\)=\Diff_k^+\(A,Q\)$.
To every d.o\. $\Delta\:P\to Q$ corresponds the homomorphism
$\varphi_{\Delta}\in\Hom_A\(P, \Diff_k^+\(Q\)\)$,
$\varphi_{\Delta}\(p\)\(a\)=\Delta\(pa\)$ under this isomorphism. The
inverse mapping takes a homomorphism $\varphi\:P\to\Diff_k^+\(Q\)$ to
an operator $\Cal D_k\com\varphi$, where $\Cal D_k\:\Diff_k^+\(Q\)\to
Q$ is a d.o\. of order $\leqslant{}k$  defined by the formula $\Cal
D_k\(\nabla\)=\nabla\(1\)$, $\nabla\in\Diff_k^+\(Q\)$.
\endroster}\endproclaim
\demo\nofrills{Proof\/\hphantom{ }} consists of a series of
automatic verifications. \QED\enddemo
This proposition has the following:
\proclaim{Corollary} {\it The commutative diagram
$$\CD
\Diff_k^+\(\Diff_l^+\(P\)\) @>\Cal D_k>> \Diff_l^+\(P\) \\
@Vc_{k,l}VV @VV\Cal D_lV \\
\Diff_{k+l}^+\(P\) @>\Cal D_{k+l}>> P
\endCD$$
uniquely defines the
operator $c_{k,l}$, which is said to be glueing operator.}\endproclaim
\definition{Definition}
A $k$-multilinear over $K$ mapping $\nabla\:A\times\ldots\times A\to
P$, $P$ being an  $A$-module, is said to be a {\it skew
multiderivation\/} if the following conditions hold: \roster
\item"{1.}" $\nabla\(a_1,\ldots,a_i,a_{i+1},\ldots,a_k\)=
-\{a_i,a_{i+1}\}\nabla\(a_1,\ldots,a_{i+1},a_i,\ldots,a_k\)$
\item"{2.}" $\nabla\(a_1,\ldots,a_{i-1},ab,a_{i+1},\ldots,a_k\)=\>
\{\nabla,a\}\{a_1\ldots
a_{i-1},a\}a\nabla\(a_1,\ldots,b,\ldots,a_k\)+\>
\{\nabla,b\}\{a_1\ldots a_{i-1}a,b\}b\nabla\(a_1,\ldots,a,\ldots,a_k\)$
\endroster\enddefinition
The set of all skew multiderivations is an $A$-module denoted by
$\Der_k\(P\)$. Obviously $\Der_1\(P\)$ is a submodule of
$\Diff_1\(A,P\)$.

{\it The Spencer $\Diff$-operator}
$S\:\Der_k\(\Diff_l^+\(P\)\)\to\Der_{k-1}\(\Diff_{l+1}^+\(P\)\)$ is
defined by the formula $$S\(\nabla\)\(a_1,\ldots,a_{k-1}\)\(a\)=
\nabla\(a_1,\ldots,a_{k-1},a\)\(1\),$$
$\nabla\in\Der_k\(\Diff^+\(P\)\)$.
\proclaim{Proposition} {\it \roster
\item $S$ is a d.o\. of order $\leqslant 1$;
\item $S^2=0$.
\endroster}\endproclaim
\demo\nofrills{Proof\/\hphantom{ }} is straightforward. \QED\enddemo
The complex $0@<<<P@<\Cal D<<\Diff^+\(P\)@<S<<\Der_1\(\Diff^+\(P\)\)
@<S<<\Der_2\(\Diff^+\(P\)\)@<<<\ldots$ is said to be {\it the Spencer
$\Diff$-complex}.
\subhead 1.3 \endsubhead
The operations $Q\mapsto\Der_k\(Q\)$, $Q\mapsto\Diff\(P,Q\)$, and
$Q\mapsto\Diff^+\(Q,P\)$ are functors from the category of all
$A$-modules $\Cal M{}od_A$ to itself. We have seen that the latter
functor is representable.  The following proposition shows that two
others functors are also representable.
\proclaim{Proposition (see \cite{43}, \cite{45}, \cite{46})}
{\it There exists a module $\Lambda^k$ \rom{(}resp\.
$\J^k\(P\)$\rom{)}),  which is called the module of $k$-form
\rom{(}resp\. $k$-jets\rom{)} over $A$, such that the  functor
$Q\mapsto\Der_k\(Q\)$ \rom{(}resp\. $Q\mapsto\Diff_k\(P,Q\)$\rom{)} is
isomorfic to the functor $Q\mapsto\Hom_A\(\Lambda^k,Q\)$  \rom{(}resp\.
$Q\mapsto\Hom_A\(\J^k\(P\),Q\)$\rom{)}.}\endproclaim
Let us denote by $j_k=j_k\(P\)\:P\to\J^k\(P\)$ the d.o\. of order
$\leqslant{}k$ that corresponds under the isomorphism
$\Diff_k\(P,\J^k\(P\)\)=\Hom_A\(\J^k\(P\), \J^k\(P\)\)$ to the
identical map $id_{\J^k\(P\)}$. Proposition 1.3 implies that the
operator $j_k$ is universal, i\.e., for every d.o\.
$\Delta\in\Diff_k\(P,Q\)$ there exists a unique homomorphism
$\psi_{\Delta}\:\J^k\(P\)\to Q$ such that
$\Delta=\psi_{\Delta}\com{}j_k$.

Now let $\Cal M$ be a subcategory of category $\Cal Mod_A$ of all
$A$-modules that  is closed under the action of the above-discussed
functors. Then it is quite natural to ask if these functors are
representable in $\Cal M$.
\example{Example} The most important example of the closed (in the
above sense) category  is that of geometrical modules $\Cal Mod_A^g$. A
module $P$ is called geometrical if $\widetilde
P=\bigcap_{\wp,k}\wp^kP=0$, where intersection is taken over all powers
of prime ideals of $A$. There is the geometrization functor $P\mapsto
P/\widetilde P$ from $\Cal Mod_A$ to $\Cal Mod_A^g$. It can easily be
checked that the functors $\Der_k$ and $\Diff_k\(P,\cdot\)$ are
representable in $\Cal Mod_A^g$, the geometrization functor
transforming the representing  objects in $\Cal Mod_A$ into the
corresponding representing objects in
$\Cal Mod_A^g$.
\endexample
\subhead 1.4 \endsubhead
A natural transformation of the functors $\Der_k$, $\Diff_k$, and their
compositions by duality gives rise to operators between the
corresponding representing objects.
\example{Example 1} The natural inclusion
$\Der_{k+l}\(P\)\to\Der_k\(\Der_l\(P\)\)$ defines the wedge product of
forms over $A$: $\Lambda^k\otimes\Lambda^l\to\Lambda^{k+l}$.
\endexample
\example{Example 2} The Spencer operator
$S\:\Der_k\(P\)\to\dot{\Der}_{k-1}\(\Diff_1^+\(P\)\)$, where the
superscribed  dot means that the $K$-module
$\Der_{k-1}\(\Diff_1^+\(P\)\)$ is supplied with $A$-module structure by
putting $a\theta=\Der_{k-1}\(\Diff_1^+\(a\)\)\theta$,
$\theta\in\Der_{k-1}\(\Diff_1^+\(P\)\)$, induces the homomorphism
$\J^1\(\Lambda^{k-1}\)\to\Lambda^k$. The composition of
$\Lambda^{k-1}@>j_1>>\J^1\(\Lambda^{k-1}\)@>>>\Lambda^k$ is called {\it
the exterior differentiation\/} operator and is denoted by
$d\:\Lambda^{k-1}\to\Lambda^k$. Using the fact that $S^2=0$, one can
easily prove that $d^2=0$. The complex
$0@>>>A@>d>>\Lambda^1@>d>>\Lambda^2@>d>>\ldots$ is said to be the  {\it
de Rham complex\/} of the algebra $A$.
\endexample
\subhead 1.5 \endsubhead
The above described algebraic formalism can be realized geometrically,
the algebra $A$ being the algebra $C^\infty\(M\)$ of smooth real
functions on a supermanifold $M$ (for the supergeometry see, e\.g.,
\cite{47}, \cite{48},  \cite{35}, \cite{49}, \cite{50}). In this
situation it may be shown  in the same way  as in non-graded case that
the standard notions of differential operator,  forms, jets, etc\.
coincide with the ones introduced above. Having this in mind,  in the
next section we shall illustrate constructions under consideration by
giving their local coordinate description.
\head 2. Adjoint Operators and Berezinian \endhead
\subhead 2.1 \endsubhead
Given an $A$-module $P$, consider the complex of homomorphisms
$$0@>>>\Diff^+\(P,A\)@>w>>\Diff^+\(P,\Lambda^1\)@>w>>
\Diff^+\(P,\Lambda^2\)@>w>>\ldots ,\tag 1 $$ where
$w\(\nabla\)=d\com\nabla\in\Diff^+\(P,\Lambda^k\)$,
$\nabla\in\Diff^+\(P,\Lambda^{k-1}\)$. Let us denote the cohomology
module in a term $\Diff^+\(P,\Lambda^n\)$ by $\ah{P}_n$, $n\geqslant
0$. Every d.o\. $\Delta\:P\to Q$ generates the natural map of the
complexes:
$$\CD
\ldots@>>>\Diff^+\(Q,\Lambda^{k-1}\)@>w>>\Diff^+\(Q,\Lambda^{k}\)
@>>>\ldots \\ @. @V\widetilde{\Delta}VV @V\widetilde{\Delta}VV @. \\
\ldots@>>>\Diff^+\(P,\Lambda^{k-1}\)@>w>>\Diff^+\(P,\Lambda^{k}\)
@>>>\ldots, \\
\endCD$$
where
$\widetilde{\Delta}\(\nabla\)=\{\Delta,\nabla\}\nabla\com\Delta\in
\Diff^+ \(P,\Lambda^k\)$, $\nabla\in\Diff^+\(Q,\Lambda^k\)$.
\definition{Definition} The operator $\Delta^*_n\:\ah{Q}_n\to\ah{P}_n$
induced by $\widetilde{\Delta}$ is called the ($n$-th) {\it adjoint
operator}.\enddefinition
Below we assume that an integer $n$ is fixed and omit the corresponding
index to simplify notation.
\proclaim{Proposition} {\it \roster
\item $\Delta^*$ has the same grading as $\Delta$.
\item If $\Delta\in\Diff_k\(P,Q\)$ then
$\Delta^*\in\Diff_k\(\ah{Q},\ah{P}\)$.
\item For all $\Delta_1\in\Diff\(P,Q\)$ and $\Delta_2\in\Diff\(Q,R\)$
we have $$\(\Delta_2\com\Delta_1\)^*=\{\Delta_2,\Delta_1\}
\Delta_1^*\com\Delta_2^*.$$
\endroster}\endproclaim
\demo{Proof} \roster
\runinitem Obvious.
\item Denote by $\[\nabla\]$ the cohomologous class of an operator
$\nabla\in\Diff^+\(P,\Lambda^n\)$, $w\(\nabla\)=0$. Then
$\Delta^*\(\[\nabla\]\)=\{\Delta,\nabla\}\[\nabla\com\Delta\]$ and we
have \> $\delta_a\(\Delta^*\)\(\[\nabla\]\)=
\{a,\Delta^*\}\Delta^*\(a\[\nabla\]\)-a\Delta^*\(\[\nabla\]\)=\>
\{a,\Delta^*\}\{a,\nabla\}\{\Delta,\nabla\com a\}\[\nabla\com
a\com\Delta\]- \{\Delta,\nabla\}
\{a,\nabla\com\Delta\}\[\nabla\com\Delta\com a\]=\>
\(a\com\Delta\)^*\(\[\nabla\]\)-\{a,\Delta\} \(\Delta\com
a\)^*\(\[\nabla\]\)=-\delta_a\(\Delta\)^*\(\[\nabla\]\),$ \>
i\.e., $\delta_a\(\Delta^*\)= -\delta_a\(\Delta\)^*$. Thus
$\delta_{a_0,\ldots,a_k}\(\Delta^*\)=\(-1\)^{k+1}
\delta_{a_0,\ldots,a_k}\(\Delta\)^*=0$. \item
$\(\Delta_2\com\Delta_1\)^*\(\[\nabla\]\)=
\{\Delta_2\com\Delta_1,\nabla\}
\[\nabla\com\Delta_2\com\Delta_1\]=\>
\{\Delta_2,\nabla\}\{\Delta_2,\Delta_1\}\Delta_1^*
\(\[\nabla\com\Delta_2\]\)=
\{\Delta_2,\Delta_1\}\Delta_1^*\(\Delta_2^*\(\[\nabla\]\)\)$.
\endroster\QED\enddemo
Let us consider some examples of adjoint operators.
\example{Example 1} Let $a\:P\to P$ be the operator of multiplication
by $a\in A$.  Then $a^*\(\[\nabla\]\)=\{a,\nabla\}\[\nabla\com
a\]=a\[\nabla\]$, i\.e., $a^*=a$.
\endexample
\example{Example 2} Let $p\:A\to P$ be the operator $p\(a\)=pa$, $a\in
A$, $p\in P$. Then $p^*\(\[\nabla\]\)=\{p,\nabla\}\[\nabla\com p\]$,
$p^*\in\Hom_A\(\ah{P},\ah{A}\)$. Thus we have a natural pairing
$\p(\cdot\,,\cdot)\:P\otimes\ah{P}\to \ah{A}$,
$\p(p,\ah{p})=p^*\(\ah{p}\)$, $\ah{p}\in\ah{P}$.
\endexample
\example{Example 3. Berezinian and Integral forms}
Let $\ldots@>>>P_{k-1}@>\Delta_k>>P_k@>>>\ldots$ be a complex of
d.o.'s. Since
$\Delta_k^*\com\Delta_{k+1}^*=\{\Delta_k,\Delta_{k+1}\}
\(\Delta_{k+1}\com\Delta_k\)^*=0$, we get a complex
$\ldots@<<<\ah{P}_{k-1}@<\Delta_k^*<<\ah{P}_k@<<<\ldots$, which is
called dual to given one. The complex dual to the de~Rham complex is
said to be the {\it complex of integral forms} and is denoted by
$$0@<<<\Sigma_0@<\delta<<\Sigma_1@<\delta<<\ldots,$$ where
$\Sigma_i=\ah{\Lambda^i}$, $\delta=d^*$. The module $\Sigma_0=\ah{A}$
is called the {\it Berezinian\/ \rom{(}or the module of  volume
forms\/\rom{)}} and will be denoted by $B$.
\endexample
The d.o.'s $\Cal D\:\Diff^+\(\Lambda^k\)\to\Lambda^k$ induce the
cohomology map: $\int\:B\to\Hm^n\(\Lambda^*\)$, so that to any element
$\omega\in B$ (volume form) corresponds the $n$-th de~Rham cohomology
class $\int\omega$. This is an algebraic version of the integration
operation. Clearly, $\int\[\nabla\]=\nabla\(1\)\mod{d\Lambda^{n-1}}$,
$\nabla\in\Diff\(A,\Lambda^n\)$.
\remark{Remark} The construction and the whole line of this subsection,
as it was mentioned in the Introduction, are principally motivated by
Penkov's explanation of the Berezin forms on a supermanifold \cite{32}.
Close approaches to the construction of the Berezin forms on a
supermanifold were suggested in \cite{51} and \cite{52}.
\endremark
\proclaim{2.2. Proposition} {\it $\int\delta\omega=0$, where
$\omega\in\Sigma_1$.}\endproclaim
\demo{Proof} Suppose $\omega=\[\nabla\]$, then
$\delta\omega=\[\nabla\com d\]$. Therefore
$\int\delta\omega=\[\nabla\com d\(1\)\]=0$.\QED\enddemo
\proclaim{2.3. Proposition (``Integration by parts'')} {\it For any
d.o\. $\Delta\:P\to Q$ and  $p\in P$, $\ah{q}\in\ah{Q}$, we have
$$\int\p(\Delta\(p\),\ah{q})=
\{\Delta,p\}\int\p(p,\Delta^*\(\ah{q}\))$$}\endproclaim
\demo{Proof} Suppose $\ah{q}=\[\nabla\]$, $\nabla\:Q\to\Lambda^n$. Then
$\int\p(\Delta\(p\),\ah{q})=
\int\{\Delta\(p\),\nabla\}\[\nabla\com\Delta\(p\)\]
\allowmathbreak=\int\{\Delta\(p\),\nabla\}\[\nabla\com\Delta\com p\]=
\int\{\Delta,\nabla\}\{\Delta,p\}\p(p,\[\nabla\com\Delta\])=
\{\Delta,p\}\int\p(p,\Delta^*\(\ah{q}\))$.\QED\enddemo
\subhead 2.4 Coordinates \endsubhead
Let $M$ be a smooth supermanifold of dimension $s|t$, $G=\Bbb Z_2$,
$K=\Bbb R \text{ or }\Bbb C$, $A=C_K^\infty\(M\)$, $P=\Gamma\(\alpha\)$
and $Q=\Gamma\(\beta\)$ the modules of smooth sections of vector
bundles over $M$. Suppose  $x=\(y_i,\xi_j\)$, $i=1,\ldots,s$,
$j=1,\ldots,t$, $x_1=y_1,\ldots,x_s=y_s$, $x_{s+1}=\xi_1,
\ldots,x_{s+t}=\xi_t$ is a coordinate system on a domain $\Cal U\subset
M$.

First of all let us compute the Berezinian, i\.e., the cohomology of
complex \thetag{1} for $P=A$.
\proclaim{Theorem (\cite{32})} {\it \roster
\runinitem $\ah{A}_n=0\qquad\text{for } n\ne s$.
\item $\ah{A}_s$ is the module of sections for the bundle of volume
forms $\Ber\(M\)$\footnote{Recall that locally sections of $\Ber\(M\)$
are written in the form $f\(x\)\text{\bf D}\(x\)$, where  $f\in
C^{\infty}\(\Cal U\)$ and {\bf D} is a basis local section that is
multiplied by the Berezin determinant of the Jacobi matrix under the
change of coordinates. The Berezin determinant of an even matrix
$\pmatrix A&B\\C&D\endpmatrix$ is equal to $\det\(A-BD^{-1}C\)\(\det
D\)^{-1}$.}.\endroster}\endproclaim
\demo{Proof} The assertion is local, so we can consider the domain
$\Cal U$ and split the complex $\Diff^+\(\Lambda^*\)$ in a tensor
product of complexes
$\Diff^+\(\Lambda^*\)_{even}\otimes\Diff^+\(\Lambda^*\)_{odd}$, where
$\Diff^+\(\Lambda^*\)_{even}$ is complex \thetag{1} on the underlying
even domain of $\Cal U$ and $\Diff^+\(\Lambda^*\)_{odd}$ is complex
\thetag{1} for the Grassmann algebra in variables
$\xi_1,\ldots,\xi_t$.

It is known that $\Hm^i\(\Diff^+\(\Lambda^*\)_{even}\)=0$ for $i\ne s$
and $\Hm^i\(\Diff^+\(\Lambda^*\)_{even}\)=\Lambda_u^s$, where
$\Lambda_u^s$ is the module of $s$-form on the underlying even domain
of $\Cal U$ (see \cite{29, section 2}). To compute the cohomology of
$\Diff^+\(\Lambda^*\)_{odd}$ consider the quotient complexes
$$0@>>>\Smbl_k\(A\)_{odd}@>>>\Smbl_{k+1}\(\Lambda^1\)_{odd}
@>>>\ldots,$$ where
$\Smbl_k\(P\)_{odd}=\Diff_k^+\(P\)_{odd}/\Diff_{k-1}^+\(P\)_{odd}$. An
easy calculation shows that these complexes are the Koszul complexes,
hence $\Hm^i\(\Diff^+\(\Lambda^*\)\)_{odd}=0$ for $i>0$ and
$\Hm^0\(\Diff^+\(\Lambda^*\)\)$ is a module of rank 1. Therefore
$\ah{A}_i=\Hm^i\(\Diff^+\(\Lambda^*\)\)=0$ for $i\ne s$ and the only
operators that represent non-trivial cocycles have the form
$f\(y,\xi\)dy_1\wedge\ldots\wedge
dy_s\frac{\partial^t}{\partial\xi_1\cdots\partial\xi_t}$.

To complete the proof it remains to check that $\ah{A}_s$ is precisely
$\Ber\(M\)$, i\.e., that changing coordinates we obtain:
$$fdy_1\wedge\ldots\wedge dy_s\frac{\partial
t}{\partial\xi_1\ldots\xi_t}=f\Ber J\(\frac{x}{z}\)dv_1\wedge\ldots
\wedge dv_s\frac{\partial t}{\partial\eta_1\ldots\eta_t}+T,$$ where
$z=\(v_i,\eta_j\)$ is a new coordinate system on $\Cal U$, $\Ber$
denotes the Berezin determinant, $J\(\frac{x}{z}\)$ is the Jacobi
matrix, $T$ is cohomologous to zero. This is an immediate consequence
of the following:
\remark{Claim} If $X=\pmatrix A&B\\C&D\endpmatrix$ and $X^{-1}=
\pmatrix
\widetilde{A}&\widetilde{B}\\\widetilde{C}&\widetilde{D}\endpmatrix$
are mutually inverse matrices written in the standard format, then
$\Ber X=\det A\cdot\det\widetilde{D}$.
\endremark
\remark{Proof} Obviously, $A\widetilde{B}+B\widetilde{D}=0$ and
$C\widetilde{B}+D \widetilde{D}=1$, whence
$D=\widetilde{D}^{-1}+CA^{-1}B$. Therefore  $$\pmatrix
A&B\\C&D\endpmatrix=\pmatrix A&B\\C&\widetilde{D}^{-1}+CA^{-1}B
\endpmatrix=\pmatrix 1&0\\CA^{-1}&1\endpmatrix\pmatrix
A&B\\0&\widetilde{D}^{-1}\endpmatrix$$ and we get $$\Ber\pmatrix
A&B\\C&D \endpmatrix=\Ber\pmatrix 1&0\\CA^{-1}&1\endpmatrix\Ber\pmatrix
A&B\\0&\widetilde{D}^{-1}\endpmatrix=\det A\cdot\det\widetilde{D}.$$
\QED\endremark\enddemo
Now let us consider the coordinate expression for adjoint operator.
Suppose $\Delta\in\Diff_k\(A,A\)$ is a scalar operator:
$\Delta=\sum_{\sigma}a_{\sigma}\frac{\partial^{|\sigma|}}{\partial
x_{\sigma}}$, where $\sigma=\(i_1,\ldots,i_r\)$ is an ordered set of
integers $1\leqslant i_j\leqslant s+t$, $|\sigma|=r$,
$\frac{\partial^{|\sigma|}}{\partial
x_{\sigma}}=\frac{\partial^{|\sigma|}}{\partial x_{i_1}\ldots\partial
x_{i_r}}$. Then we have $$\Delta^*=\sum_{\sigma}\(-1\)^{\Theta}\frac
{\partial^{|\sigma|}}{\partial x_{\sigma}}\com a_{\sigma},$$ where
$\Theta=|\sigma|+\widetilde{a_{\sigma}}\sum_{j=1}^{|\sigma|}
\widetilde{x_{i_j}}+\sum_{1\leqslant
j<k\leqslant|\sigma|}\widetilde{x_{i_j}}\widetilde{x_{i_k}}$ and  tilde
over an object denotes the parity of the object. For a matrix d.o\.
$\Delta=\|\Delta_{ij}\|$ one has
$\(\Delta^*\)_{ij}=\(-1\)^{\widetilde{i}\widetilde{j}}
\(\Delta_{ji}\)^*$.
\head 3. Spencer Complexes and Green's Formula \endhead
\subhead 3.1 \endsubhead
{}From now onwards we assume that the module $\Lambda^1$ is of finite
type and projective. This implies that the same is true for the modules
$\Lambda^k$ and $\J^k\(A\)$ and, also, that for any projective module
the Spencer  $\Diff$-complex is exact (see\cite{45}).  In this case it
is expedient to consider another variant of construction of adjoint
operator.

Set $\ao{P}=\Hom\(P,B\)$. Obviously,
$\Sigma_i=\ah{\Lambda^i}=\ao{\(\Lambda^i\)}=\Der_i\(B\)$,
$\ah{\J^k\(A\)}=\ao{\J^k\(A\)}=\Diff_k\(A,B\)$.
\definition{Definition} For an operator $\Delta\in\Diff_k\(A,B\)$ we
define the operator $\ao{\Delta}\:A\to B$,
$\ao{\Delta}\(a\)=\{\Delta,a\}j_k^*\(a\Delta\)$, $a\in A$, which, for
economy of language, will be called {\it adjoint}.
\enddefinition
\proclaim{Proposition} {\it \roster
\runinitem The grading of $\ao{\Delta}$ is equal to the one of
$\Delta$.
\item If $\Delta\in\Diff_k\(A,B\)$ then $\ao{\Delta}\in\Diff_k\(A,B\)$.
\item $\ao{\omega}=\omega$, $\omega\in B=\Diff_0\(A,B\)$.
\item If $X\in\Der_1\(B\)$ then $X+\ao{X}\in\Diff_0\(A,B\)=B$ and
$X+\ao{X}=\delta\(X\)$.
\item $\ao{\(a\Delta\)}=\{a,\Delta\}\ao{\Delta}\com a$,
$\Delta\in\Diff\(A,B\)$, $a\in A$.
\item If $\Delta\(a\)=\[\nabla_a\]$, $\Delta\in\Diff\(A,B\)$, $a\in A$,
$\nabla_a\in\Diff^+ \(\Lambda^n\)$, then
$\ao{\Delta}\(a\)=\[\square_a\]$, where
$\square_a\(a'\)=\{a,a'\}\nabla_{a'}\(a\)$. \item
$\ao{\(\ao{\Delta}\)}=\Delta$.
\endroster}\endproclaim
\demo{Proof}\roster
\runinitem Obvious.
\item
$\delta_a\(\ao{\Delta}\)\(a'\)=\{a,\ao{\Delta}\}\ao{\Delta}\(aa'\)-
a\ao{\Delta}\(a'\)=\>
\{\ao{\Delta},a'\}j_k^*\(aa'\Delta\)-
\{\ao{\Delta},a'\}aj_k^*\(a'\Delta\)=\>
\{\ao{\Delta},a'\} \delta_a\(j_k^*\)\(a'\Delta\)$, so
$\delta_{a_0,\ldots,a_k}\(\ao{\Delta}\)\(a'\)=\>
\{\ao{\Delta},a'\}\delta_{a_0,\ldots,a_k}\(j_k^*\)\(a'\Delta\)=0$.
\item Obvious.
\item Clearly, $\delta_a\(j_1\)=j_1\(a\)-aj_1\(1\)\in\J^1\(A\)$, hence
$$\(\delta_a\(j_1\)\)^*\(\Delta\)=\{a,\Delta\}\Delta\(a\)-a\Delta\(1\)=
\delta_a\(\Delta\)\(1\),
\qquad\Delta\in\Diff_1\(A,A\).$$
Thus
$\delta_a\(X+\ao{X}\)\(1\)=\delta_a\(X\)\(1\)+\delta_a\(j_1^*\)\(X\)=
\delta_a\(X\)\(1\)-\delta_a\(j_1\)^*\(X\)\allowmathbreak=0$. Further,
$\delta\(X\)=j_1^*\(\gamma^*\(X\)\)$, where
$\gamma\:\J^1\(A\)\to\Lambda^1$ is the natural projection. Obviously,
$\gamma^*\:\Der_1\(B\)\to\Diff_1\(A,B\)$ is the natural inclusion, so
that $\delta\(X\)=\ao{X}\(1\)=X+\ao{X}$.
\item
$\ao{a\Delta}\(a'\)=\{a\Delta,a'\}j_k^*\(a'a\Delta\)=
\{a\Delta,a'\}\{a'a,
\Delta\} \ao{\Delta}\(a'a\)=\>\{a,\Delta\}\(\ao{\Delta}\com a\)\(a'\)$.
\item Obvious.
\item Let $\Delta\(a\)=\[\nabla_a\]$ and
$\ao{\Delta}\(a\)=\[\square_a\]$. Then
$\ao{\(\ao{\Delta}\)}\(a\)=\[\widetilde{\square_a}\]$, where
$\widetilde{\square_a}\(a'\)=\{a,a'\}\square_{a'}\(a\)=\nabla_a\(a'\)$,
i\.e.,  $\ao{\(\ao{\Delta}\)}\(a\)=\[\nabla_a\]=\Delta\(a\)$.
\endroster\QED\enddemo
Now let us define the adjoint operator $\ao{\Delta}$ in the general
case when $\Delta\in\Diff\(P,\ao{Q}\)$, the $P$ and $Q$ being
$A$-modules. As in non-graded case consider the family of operators
$\Delta\(p,q\)\in\Diff\(A,B\)$, $p\in P$, $q\in Q$,
$\Delta\(p,q\)\(a\)=\{p,a\}\{q,a\}\p(\Delta\(ap\),q)$, $a\in A$; the
brackets $\p(\cdot\,,\cdot)$ denote the natural pairing $\ao{Q}\otimes
Q\to B$. Then set
$\p(\ao{\Delta}\(q\),p)=\{q,p\}\ao{\Delta\(p,q\)}\(1\)$,
$\ao{\Delta}\in\Diff\(Q,\ao{P}\)$.
\proclaim{Proposition} {\it \roster
\runinitem For any $\Delta\in\Diff_k\(P,\ao{Q}\)$ the operator
$\ao{\Delta}$ is well-defined and of order $\leqslant{}k$.
\item For every $\Delta\in\Diff\(P,\ao{Q}\)$ we have
$\ao{\(\ao{\Delta}\)}=\Delta$.
\item If the modules $P$, $Q$, $\ao{Q}$ are projective, then for
$\Delta\in\Diff\(P,\ao{Q}\)$ the adjoint operator
$\ao{\Delta}\:Q\to\ao{P}$ coincides with the composition of
$Q@>>>\ao{\ao{Q}}@>\Delta^*>>\ao{P}$, where $Q\to\ao{\ao{Q}}$ is the
natural homomorphism.
\endroster}\endproclaim
\demo{Proof} Statements \therosteritem1 and \therosteritem2 are
straightforward.
\roster\item[3] Take $\Delta\:P\to\ao{Q}$, $p\in P$, $q\in Q$. We have
$\Delta=\psi_{\Delta}j_k$, so
$\p(\Delta^*\(q\),p)=\p(j_k^*\psi_{\Delta}^*\(q\),p)=
j_k^*\(\psi_{\Delta}^*\(q\)\com p\)=\ao{\(\psi_{\Delta}^*\(q\)\com
p\)}\(1\)= \{q,p\}\ao{\Delta\(p,q\)}\(1\)$
\endroster\QED\enddemo
Now let us consider one more example of adjoint operators.
\subhead 3.2 Example: The Spencer Operators \endsubhead
The complex $$0@>>>P@>j>>\JI\(P\)@>s>>\JI\(P\)\otimes\Lambda^1@>s>>
\JI\(P\)\otimes\Lambda^2@>s>>\ldots,$$ where
$s\(j\(p\)\otimes\omega\)=j\(p\)\otimes d\omega$, is
called the Spencer $\J$-complex (for details see \cite{45}). Note that
$s$ is a d.o\. of order $\leqslant{}1$. Let us compute the dual
complex. Clearly, for the projective module $P$ we have
$\ah{\JI\(P\)\otimes\Lambda^i}= \ao{\(\JI\(P\)\otimes\Lambda^i\)}=
\Der_i\(\Hom\(\JI\(P\),B\)\)=\Der_i\(\Diff\(P,B\)\)=
\Der_i\(\Diff^+\(\ao{P}
\)\)$. To describe $s^*$ we need the following remark.

Let $\Delta\(P\)\:\F\(P\)\to\G\(P\)$ be a {\it natural\/} d.o., $\F$
and $\G$ be  functors on a category of projective modules over $A$. For
any functors $\F$ denote by $\dot{\F}\(P\)$ the abelian group $\F\(P\)$
supplied with $A$-module structure induced by the $A$-module structure
in $P$. Then $\Delta$ generates the natural homomorphism
$\dot{\Delta}\:\dot{\F}\to\dot{\G}$.
\proclaim{Lemma} $\(\Delta^*\)^{\tsize\cdot}=\(\dot{\Delta}\)^*$
\endproclaim
\demo\nofrills{Proof\/\hphantom{ }} is trivial. \QED\enddemo
The lemma immediately implies that the complex dual to the Spencer
$\J$-complex for module $P$ is the Spencer $\Diff$-complex for module
$\ao{P}$. On the other hand we have
$\ah{\JI\(P\)\otimes\Lambda^k}=\Hom\(\JI\(P\),\Der_k\(B\)\)=
\Diff\(P,\Sigma_k\)$ and the operator $s^*\:\Diff\(P,\Sigma_k\)\to
\Diff\(P,\Sigma_{k-1}\)$ has the form $s^*\(\nabla\)=\delta\com\nabla$,
$\nabla\in\Diff\(P,\Sigma_k\)$; $j^*\(\nabla\)=\ao{\nabla}\(1\)$.
Bringing all this together we can state the following:
\proclaim{Theorem} {\it For a projective module $P$ there is an
isomorphism of complexes:
$$\minCDarrowwidth{0.7true cm}\CD
0@<<<\ao{P}@<\Cal D<<\Diff^+\(\ao{P}\)@<S<<\Der_1\(\Diff^+\(\ao{P}\)\)
@<S<<\Der_2\(\Diff^+\(\ao{P}\)\)@<<<\ldots \\
@. @| @V\psi VV @V\psi VV @V\psi VV @. \\
0@<<<\ao{P}@<\mu<<\Diff\(P,B\)@<\omega<<\Diff\(P,\Sigma_1\)@<\omega<<
\Diff\(P,\Sigma_2\)
@<<<\ldots,
\endCD$$
where $\omega\(\nabla\)=\delta\com\nabla$,
$\nabla\in\Diff\(P,\Sigma_k\)$, $\mu\(\nabla\)=\ao{\nabla}\(1\)$, the
maps $\psi$ are the composition of isomorphisms:
$$\Der_k\(\Diff^+\(\ao{P}\)\)@>\Der_k\(\com\)>>\Der_k\(\Diff\(P,B\)\)
@>>>\Diff\(P,\Der_k\(B\)\)@>>>\Diff\(P,\Sigma_k\).$$
}\endproclaim
This theorem has the following:
\proclaim{Corollary (Green's Formula)} {\it If the Spencer
$\Diff$-cohomology of the Berezinian $B$ in the term $\Diff^+\(B\)$ is
trivial, then for any $\Delta\in\Diff\(P,\ao{Q}\)$, $p\in P$, $q\in Q$,
we have $$\p(\Delta\(p\),q)- \{\Delta,p\}\p(p,\ao{\Delta}\(q\))=\delta
G,$$ where $G\in\Sigma_1$  is an integral 1-form.}\endproclaim
\demo{Proof} Suppose $\nabla\in\Diff^+\(B\)$; then
$\(\nabla-\nabla\(1\)\)\in\ker\Cal D$, hence there exists
$\square\in\Der_1\(\Diff^+\(B\)\)$ such that
$S\(\square\)=\nabla-\nabla\(1\)$. It follows from the theorem that
$\nabla^*\(1\)-\nabla\(1\)=\psi\com
S\(\square\)=\omega\(\psi\(\square\)\)$. Therefore
$\nabla^*\(1\)-\nabla\(1\)=\delta G$, where $G=\psi\(\square\)\(1\)$.
Letting $\nabla=\Delta\(p,q\)$, we get the Green formula. \QED\enddemo
\remark{Remark}
If $B$ is projective then there exists an $A$-homomorphism
$\varkappa\:\ker\Cal D\to \Der_1\(\Diff^+\(B\)\)$, such that
$S\com\varkappa=id$, and we can take  $\square=\varkappa\(
\Delta-\Delta\(1\)\)$ and, therefore, $G=G_{\varkappa}=
\psi\(\varkappa\(\Delta\(p,q\)-\p(\Delta\(p\),q) \)\)\(1\)$. Since the
Spencer complex of $B$ is exact, for any two  homomorphisms $\varkappa$
and $\varkappa'$ one can find a homomorphism $f\:\ker\Cal
D\to\Der_2\(\Diff^+\(B\)\)$  such that $\varkappa-\varkappa'=S\com f$.
Hence  $G_{\varkappa}-G_{\varkappa'}=\delta F$, where $F=\psi\(f\(
\Delta\(p,q\)-\p(\Delta\(p\),q)\)\)\(1\)$.
\endremark
\subhead 3.3 \endsubhead
In this subsection we describe a spectral sequence, which establishes
the relationship between the de~Rham cohomology and the homology of the
complex of integral forms.
\proclaim{Proposition (Poincar\'e duality)} {\it There is a spectral
sequence, $\(E^r_{*,*},d^r_*\)$, with
$$E^2_{p,q}=\Hm_p\(\(\Sigma_*\)_{-q}\),$$ the homology of complexes of
integral forms, and converging to the de~Rham cohomology
$\Hm\(\Lambda^*\)$.}\endproclaim
\demo{Proof} Let $K_{p,q}=\Der_p\(\Diff^+\(\Lambda^{-q}\)\)$, $d'$ be
the Spencer operator $d'\:K_{p,q}\to K_{p-1,q}$, and
$d''=\(-1\)^p\Der_p\(\Diff^+\(d\)\)$, $d''\:K_{p,q}\to K_{p,q-1}$. Then
$\{K_{*,*},d',d''\}$ is a double complex. Obviously,
$\Hm^I\(K\)=\Hm\(K_{*,*},d'\)=\Lambda^{-q}$ for $p=0$ and 0 for
$p\ne0$. Therefore in the second spectral sequence
${}^{II}E^2_{p,q}={}^{II}\Hm_{p,q}
\({}^I\Hm\(K\)\)=\Hm_{p,q}\({}^I\Hm_{*,*}\(K\),d''\)=
\Hm^{-q}\(\Lambda^*\)$
for $p=0$ and ${}^{II}E^2_{p,q}=0$ for $p\ne0$. Thus the second
spectral sequence converge to the de~Rham cohomology. From the first
spectral sequence we get
${}^IE^2_{p,q}={}^I\Hm_{p,q}\({}^{II}\Hm\(K\)\)=
{}^I\Hm_{p,q}\(\Hm\(K_{*,*},d''\)\)=\Hm_{p,q}\(\Der_*\(B\),d'\)=
\Hm_p\(\(\Sigma_*\)_{-q}\)$.\QED\enddemo
\proclaim{Corollary} {\it If $\ah{A_i}=0$ for all $i\ne n$, then
$\Hm_i\(\Sigma_*\)=\Hm^{n-i}\(\Lambda^*\)$.}\endproclaim
\head 4. Quadratic Lagrangians, the Euler Operator,
and the Noether theorem \endhead
In this section $P$ and $Q$ are projective modules.
\subhead 4.1 \endsubhead
Consider the complex $$0@<<<\Diff^{sym}\(P,\ao{P}\)@<\mu<<
\Diff^{sym}_{\(2\)}\(P,B\)@<\omega<<\Diff^{sym}_{\(2\)}\(P,\Sigma_1\)
@<\omega<<\ldots,\tag 2 $$ where  $\Diff^{sym}_{\(2\)}\(P,Q\)$ denotes
the submodule of $\Diff\(P, \Diff\(P,Q\)\)$ consisting of all symmetric
bidifferential $Q$-valued operators
$\nabla\(p_1\)\(p_2\)=\{p_1,p_2\}\allowmathbreak\nabla\(p_2\)\(p_1\)$,
$\nabla\in\Diff\(P,\Diff\(P,Q\)\)$, $\Diff^{sym}\(P,\ao{P}\)$ is the
module of all self-adjoint operators from $\Diff\(P,\ao{P}\)$,
$\omega\(\nabla\)=\delta\com\nabla$,
$\nabla\in\Diff^{sym}_{\(2\)}\(P,\Sigma_k\)$, $k>0$, and
$\mu\(\nabla\)\(p\)= \ao{\(\nabla\(p\)\)}\(1\)$, $p\in P$.
\proclaim{Theorem} {\it This complex is acyclic.}\endproclaim
\demo{Proof} From theorem 3.2 it follows easily that the complex
$$0@<<<\Diff\(P,\ao{P}\)@<\widetilde{\mu}<<\Diff\(P,\Diff\(P,B\)\)
@<\widetilde{\omega}<<
\Diff\(P,\Diff\(P,\Sigma_1\)\)@<\widetilde{\omega}<<\ldots,\tag 3 $$
where  $\widetilde{\omega}\(\nabla\)=\delta\com\nabla$,
$\widetilde{\mu}\(\nabla\)\(p\)= \ao{\(\nabla\(p\)\)}\(1\)$, is
acyclic.  To prove the theorem it is sufficient to show that this
complex split into the symmetric and the skew-symmetric parts. To do
this let us check that the involution $\rho$ of complex \thetag{3},
$\rho\(\nabla\)\(p_1\)\(p_2\)=\{p_1,p_2\} \nabla\(p_2\)\(p_1\)$,
$\nabla\in\Diff\(P,\Diff\(P,\Sigma_k\)\)$ and $\rho\(\nabla\)=
\ao{\nabla}$, $\nabla\in\Diff\(P,\ao{P}\)$, is an authormorphism of
this complex. The fact that
$\widetilde{\omega}\com\rho=\rho\com\widetilde{\omega}$ is obvious. Let
us verify that $\widetilde{\mu}\com\rho=\rho\com\widetilde{\mu}$. Take
$\Delta\in\Diff\(P,\Diff\(P,B\)\)$ and let
$\delta\(p_1\)\(p_2\)=\[\nabla_{p_1,p_2}\]$,
$\nabla_{p_1,p_2}\in\Diff^+\(\Lambda^n\)$. It follows from proposition
3.1 that $\p(\widetilde{\mu}\(\Delta\)\(p_1\),p_2)=\[\square\]$, where
$\square\in\Diff^+\(\Lambda^n\)$,
$\square\(a\)=\{p_2,a\}\nabla_{p_1,ap_2}\(1\)$, $p_1,p_2\in P$.
Therefore
$\p(\rho\widetilde{\mu}\(\Delta\)\(p_1\),p_2)=\p(\ao{\widetilde{\mu}
\(\Delta\)}\(p_1\),p_2)=\{p_1,p_2\}\[\square'\]$, where
$\square'\(a\)=\{p_1,a\}\{p_2,a\}\nabla_{ap_2,p_1}\(1\)$. On the other
hand $\p(\widetilde{\mu}\rho\(\Delta\)\(p_1\),p_2)=\[\square''\]$,
where $\square''\(a\)=\{p_2,a\}\nabla'_{p_1,ap_2}\(1\)$,
$\p(\rho\(\Delta\)\(p_1\),p_2)=\[\nabla'_{p_1,p_2}\]$. Clearly,
$\nabla'_{p_1,p_2}= \{p_1,p_2\}\nabla_{p_2,p_1}$, hence
$\square''\(a\)=\{p_2,a\}\{p_1,ap_2\}\nabla_{ap_2,p_1}\(1\)$.
\QED\enddemo
\subhead 4.2 Lagrangian Formalism \endsubhead
\definition{Definition} The space $\Cal L ag\(P\)$ of {\it quadratic
Lagrangians\/} on $P$ is defined as the cokernel of the operator
$\omega\:\Diff^{sym}_{\(2\)}\(P,\Sigma_1\)\to
\Diff^{sym}_{\(2\)}\(P,B\)$. An operator
$L\in\Diff^{sym}_{\(2\)}\(P,B\)$ is called the {\it density\/} of
quadratic Lagrangian $\Cal L$ if $\Cal L=L\mod\im\omega$.
\enddefinition
{}From theorem 4.1 we see that the operator $\mu$ gives rise to an
isomorphism of $\Cal L ag\(P\)$ to the submodule of the module
$\Diff\(P,\ao{P}\)$ consisting of self-adjoint operators. This
isomorphism is said to be the {\it Euler operator} and denoted by
$\Cal E$.
\subhead 4.3 Conservation laws \endsubhead
Let $\Delta\in\Diff_k\(P,Q\)$ and $E=\{p\in P|\Delta\(p\)=0\}$ is the
corresponding equation. The operator $\Delta$ generates the chain map
$\Omega_{\Delta}$ of the complexes \thetag{2}:
$$\CD
0@<<<\Diff\(Q,B\)@<\omega<<\Diff\(Q,\Sigma_1\)
@<\omega<<\Diff\(Q,\Sigma_2\)@<\omega<<\ldots\\
@. @VV\Omega_{\Delta}V @VV\Omega_{\Delta}V @VV\Omega_{\Delta}V @. \\
0@<<<\Diff\(P,B\)@<\omega<<\Diff\(P,\Sigma_1\)
@<\omega<<\Diff\(P,\Sigma_2\)@<\omega<<\ldots
\endCD$$

{\it {\rm(}Linear{\rm)} conservation laws\/} for the equation $E$ are
defined  by analogy with non@-graded case (see \cite{29}) as classes of
1-dimensional homology of the complex $\coker\Omega_{\Delta}$. Let us
denote the group of linear conservation laws for the equation
$\Delta=0$ by $\Cl\(\Delta\)=\Hm_1\(\coker\Omega_{\Delta}\)$. The
following theorem and the corollary describe the group $\Cl\(\Delta\)$.
\proclaim{Theorem} {\it There exists an exact sequence
$$0@>>>\Hm_1\(\im\Omega_{\Delta}\)@>>>
\Hm_0\(\ker\Omega_{\Delta}\)@>>>\ker\Delta^*@>>>\Cl\(\Delta\)@>>>0.$$
}\endproclaim
\demo{Proof} It follows from theorem 4.1 that exact homology sequences
corresponding to the short exact sequences of complexes
$$0@>>>\ker\Omega_{\Delta}@>>>\Diff\(Q,\Sigma_*\)@>>>\im\Omega_{\Delta}
@>>>0$$
$$0@>>>\im\Omega_{\Delta}@>>>\Diff\(p,\Sigma_*\)
@>>>\coker\Omega_{\Delta}@>>>0$$ have the form
$$0@>>>\Hm_1\(\im\Omega_{\Delta}\)@>>>\Hm_0\(\ker\Omega_{\Delta}\)
@>i_1>>\ao{Q}@>i>>\Hm_0
\(\im\Omega_{\Delta}\)@>>>0$$
$$0@>>>\Hm_1\(\coker\Omega_{\Delta}\)@>>>\Hm_0\(\im\Omega_{\Delta}\)
@>j>>\ao{P}@>>>\Hm_0
\(\coker\Omega_{\Delta}\)@>>>0.$$
It is straightforward to check that the composition $j\com
i\:\ao{Q}\to\ao{P}$ coincides with the adjoint operator
$\Delta^*\:\ao{Q}\to\ao{P}$. Hence $\ker \Delta^*/\im i_1$ is
isomorphic to $\ker j=\Cl\(\Delta\)$ and we get the desired exact
sequence.\QED\enddemo
\proclaim{Corollary} {\it If $\ker\Omega_{\Delta}=0$ then the group of
linear conservation laws $\Cl\(\Delta\)$ is isomorphic to
$\ker\Delta^*$.}\endproclaim
Let us give an explicit expression for the map
$\ker\Delta^*\to\Cl\(\Delta\)$. Suppose
$\ao{q}\in\ker\Delta^*\subset\ao{Q}$. Then, choosing a homomorphism
$\varkappa$ (see Remark 3.2), we have a d.o\. from $P$ to $\Sigma_1$,
$p\mapsto G_{\varkappa}\(\Delta\(p,\ao{q}\)\)$. The Green formula
yields $\p(\Delta\(p\),\ao{q})=\delta
G_{\varkappa}\(\Delta\(p,\ao{q}\)\)$. Hence the operator $p\mapsto
G_{\varkappa}\(\Delta\(p,\ao{q}\)\)$ is an 1-cocycle of the complex
$\coker\Omega_{\Delta}$ and we obtain a map
$\chi\:\ker\Delta^*\to\Cl\(\Delta\)$, where $\chi\(\ao{q}\)$ is the
conservation law corresponding to the operator $p\mapsto
G_{\varkappa}\(\Delta\(p,\ao{q}\)\)$. Let us show that this is the map
under  consideration. If $\nabla$ is an 1-cocycle of
$\coker\Omega_{\Delta}$, the element $\ao{q}$ from $\ker\Delta^*$
corresponding to it according to the proof of theorem 4.3 can be found
as $\ao{q}=\mu\(\square\)$, where $\square\in\Diff\(Q,B\)$ satisfy the
relation $\square\com\Delta=\delta\com\nabla$. If $\nabla$ is the
operator $p\mapsto G_{\varkappa}\(\Delta\(p,\ao{q}\)\)$ then
$\square=\ao{q}\in\Diff\( P,B\)$ and
$\mu\(\square\)=\mu\(\ao{q}\)=\ao{q}$.
\subhead 4.4 The Noether Theorem \endsubhead
We start with a description of transformations of the objects that
Noether's theorem includes.

Let $\DER\(P\)=\{\(X_P,X\)|\,X\in\Der_1\(A\), X_P\in\Diff_1\(P,P\),
\text{and}  \,\,\forall a\in A, \forall p\in P \quad
X_P\(ap\)=\{X_P,a\}aX_P\(p\)+X\(a\)p\, \}$. If $X\in\Der_1\(A\)$ then
$\(X_B,X\)\in\DER\(B\)$, where $X_B=-X^*$ (more generally, if
$\(X_P,X\)\in\DER\(P\)$ then one can define
$\(X_{\ao{P}},X\)\in\DER\(\ao{P}\)$, $X_{\ao{P}}=-\(X_P\)^*$). Given
$\(X_P,X\)\in \DER\(P\)$, define
$\(X_{\Diff},X\)\in\DER\(\Diff\(P,B\)\)$ by the formula
$X_{\Diff}\(\Delta\)= \[X,\Delta\]=
X_{B}\com\Delta-\{X,\Delta\}\Delta\com X_P$. For
$L\in\Diff\(P,\Diff\(P,B \)\)$ we put $X_P\(L\)=X_{\Diff}\com
L-\{X,L\}L\com\allowmathbreak X_P$. If $L\in\Diff^{sym}_{\(2\)}\(P,B\)$
then $X_P\(L\)\in\Diff^{sym}_{\(2\)}\(P,B\)$ and is called the {\it
variation\/} of $L$ under the infinitesimal transformation $X_P$.
Clearly, $X_P$ generates the map of Lagrangians on $P$: $X_P\:\Cal L
ag\(P\) \to\Cal L ag\(P\)$.

{}From the definition of variation it follows that \newline
$X_P\(L\)\(p_1\)
\(p_2\)=X_B\(L\(p_1\)\(p_2\)\)-\{X,L\}\{X,p_1\}L\(p_1\)
\(X_P\(p_2\)\)-\> \{X,L\}L\(X_P\(p_1\)\)\(p_2\)$.
\newline Further, using proposition 3.1\therosteritem4 we get
$$X_B\(L\(p_1\)\(p_2\)\)=
-\{X,L\}\{X,p_1\}\{X,p_2\}\delta\(L\(p_1\)\(p_2\)\com X\).$$
It follows from Green's formula that $\forall p_1,p_2\in P$
$$L\(p_1,p_2\)= \p(\Cal E_L\(p_1\),p_2)+\delta
G_{\varkappa}\(L\(p_1\)\(p_2,1\)\).$$ Combining these  formulae, we
obtain the following {\it formula for the first variation\/}:\newline
$X_P\(L\) \(p_1\)\(p_2\)=-\{L,p_2\}\{p_1,p_2\}\p(X_P\(p_2\),\Cal
E_L\(p_1\))-\> \{L,p_1\}\p(X_P\(p_1\),\Cal E\(p_2\))-\delta
n_{\varkappa}\(p_1,p_2\),$  \newline where
$n_{\varkappa}\(p_1,p_2\)=\>\{X,L\}\{X,p_1\}\{X,p_2\}L\(p_1\)\(p_2\)
\com X+\{X,L\}\{X,p_1\}G_{\varkappa}\(p_1,X_P\(p_2\)\)+\>
\{X,L\}\{X,p_2\}\{p_1,p_2\}G_{\varkappa}\(p_2,X_P\(p_1\)\)$.
\definition{Definition} $X_P\in\DER\(P\)$ is said to be a {\it
symmetry\/} of a Lagrangian $\Cal L$ if \newline $X_P\(\Cal L\)=0$.
\enddefinition
{}From theorem 4.1 it follows that a symmetry of $\Cal L$ is a symmetry
of the operator $\Cal E_{\Cal L}=\Cal E\(\Cal L\)$, i\.e., $X_P\(\Cal
E_{\Cal L}\)=X_{\ao{P}} \com\Cal E_{\Cal L}- \{X,\Cal E_{\Cal L}\}\Cal
E_{\Cal L}\com X_P=0$, and conversely.

If $X_P$ is a symmetry of $\Cal L$, then $X_P\(L\)=\omega\(L'\)$, where
$L$ is a density of $\Cal L$, $L'\in\Diff^{sym}_{\(2\)}\(P,\Sigma_1\)$.
Consider the integral 1-form
$n_{\varkappa}\(p_1,p_2\)+L'\(p_1\)\(p_2\)$. The first variation
formula implies that this integral form is closed whenever $p_1,
p_2\in\ker\Cal E_{\Cal L}$. Clearly, its homological class does not
depend on the choice of $\varkappa$ and $L'$. Thus we have proved the
following:
\proclaim{Theorem (Noether)} {\it If $X_P$ is a symmetry of the
Lagrangian  $\Cal L=L\mod\im\omega$ and the module $P$ is projective,
then the map $p\mapsto n_{\varkappa}\(p,p\)+L'\(p\) \(p\)$, $p\in P$,
gives rise to a conservation law of the equation $\Cal E_{\Cal
L}=0$.}\endproclaim
\head Appendix: Right Connections \endhead
This appendix provides a lightening sketch of the algebraic formalism
of right connections.
\remark{Remark} This issue is also treated in \cite{53}, the approach,
although different, being close to the approach suggested here.
\endremark
\subhead 1 \endsubhead
We begin with a collection of a few facts about (usual) linear
connections.

One can  consider a {\it connection\/} on an $A$-module $P$ as an
$A$-homomorphism of grading zero $\gamma\:P\mapsto\J^1\(P\)$ satisfying
$\nu\com\gamma=id_P$, where $\nu\:\J^1\(P\)\to P$ is the natural
projection.

The composition $\Lambda^i\otimes P@>id\otimes\gamma>>
\Lambda^i\otimes\J^1\(P\)@>s>>\Lambda^{i+1}\otimes P$ of
$\gamma$ and the Spencer operator $s$ is called the {\it de~Rham
operator\/} associated with $\gamma$ and denoted by
$d\:\Lambda^i\otimes P\to\Lambda^{i+1}\otimes P$. The first de~Rham
operator $d\:P\to\Lambda^1\otimes P$ gives rise in a natural way to a
{\it covariant derivative}, i\.e., a homomorphism
$\nabla\:\Der_1\(A\)\to\Diff_1\(P,P\)$, $X\mapsto\nabla_X$, such that
$\nabla_X\(ap\)=\{X,a\}a\nabla_X\(p\)+X\(a\)p$, $X\in\Der_1\(A\)$. The
de~Rham sequence with values in $P$ $$0@>>>P@>d>>\Lambda^1 \otimes
P@>d>>\Lambda^2\otimes P@>d>>\ldots$$ is not a complex in general. The
operator $d^2\:\Lambda^i\otimes P\to\Lambda^{i+2}\otimes P$ is said to
be  {\it curvature\/} of the connection $\gamma$. It is straightforward
to check that $d^2$ is a $\Lambda^*$-linear operator and, therefore,
for $P$ projective the curvature $d^2$ is multiplication by a
$R_{\gamma}\in\Lambda^2\otimes\Hom_A\(P,P\)$.
\subhead 2 \endsubhead
We define a right connection in a dual way.
\definition{Definition} A {\it right connection\/} on $P$ is defined to
be an $A$-homomorphism of grading zero $\lambda\:\Diff^+_1\(P\)\to P$
satisfying $\lambda\com\iota=id_P$, where $\iota\:P\to\Diff^+_1\(P\)$
is the natural inclusion.
\enddefinition
Given a module $P$ with a right connection $\lambda$, one can carry out
the construction of the {\it sequence of integral form with values in}
$P$ by letting the operator $\delta\:\Der_{i+1}\(P\)\to\Der_i\(P\)$ be
the composition
$\Der_{i+1}\(P\)@>S>>\Der_i\(\Diff^+_1\(P\)\)
@>\Der_i\(\lambda\)>>\Der_i\(P\)$
of the Spencer operator $S$ and $\Der_i\(\lambda\)$:
$$0@<<<P@<\delta<<\Der_1\(P\)@<\delta<< \Der_2\(P\)@<\delta<<\ldots$$

The first operator $\delta\:\Der_1\(P\)\to P$ provides a {\it right
covariant derivative}, i\.e., a homomorphism
$\nabla\:\Der_1\(A\)\to\Diff^+_1\(P,P\)$, $X\mapsto\nabla_X$, such that
$$\nabla_X\(ap\)=\{X,a\}a\nabla_X\(p\)-X\(a\)p,\qquad X\in \Der_1\(A\).
\tag 1 $$
\remark{Remark} This ``right Leibniz rule'' may be reformulated by the
following way. The operator $\nabla$ can be extended to an
$A$-homomorphism $\nabla\:\Diff_1\(A,A\)\to\Diff^+_1\(P,P\)$ by putting
$\nabla_{id_A}=id_P$. Then \thetag{1} means that the map
$\nabla\:\Diff^+_1\(A,A\)\allowmathbreak \to\Diff_1\(P,P\)$ is also an
$A$-homomorphism.
\endremark
The operator $\delta^2\:\Der_{i+2}\(P\)\to\Der_i\(P\)$ is called the
{\it curvature\/} of the right connection $\lambda$. A direct
calculation shows that this is a $\Der_*\(A\)$-linear operator and, so,
for $P$ projective the curvature $\delta^2$ can be interpreted as inner
product with a $R_{\lambda}\in\Lambda^2\otimes\Hom_A\(P,P\)$.
\example{Example} If on a projective module $P$ there is a connection
$\gamma\:P\to\J^1\(P\)$, then on $\ao{P}$ there is a right connection
$\lambda=\gamma^*\: \ao{\(\J^1\(P\)\)}=\Diff^+_1\(\ao{P}\)\to\ao{P}$.
In particular, the obvious flat connection on $A$
$\gamma\:A\to\J^1\(A\)$, $\gamma\(a\)=aj_1\(1\)$, provides the
canonical flat right connection on the Berezinian $B$. The complex of
integral forms with values in $\ao{P}$ is dual to the de~Rham complex
with coefficients in $P$.
\endexample
\head Acknowledgments \endhead
The author is grateful to Professor A. M. Vinogradov for very
stimulating attention to this work and to I. S. Krasil$'$shchik, M. M.
Vinogradov, and D. M. Guessler for useful discussions and their remarks
on the preprint of this work.

He also wishes to thank the SISSA for warm hospitality and pleasant
atmosphere  in which this work was done and the Italian Ministero degli
Affari Esteri  for a fellowship.

It is a great pleasure for the author to give his special thanks to the
scientific attach\'e at the Italian Embassy in Moscow Professor G.
Piragino for an understanding and  help which made possible his stay at
the SISSA.

Finally, many thanks to the referee who indicated some of the relevant
references added in the text.
\enddocument